\documentclass{aa}  

\usepackage{graphicx}
\usepackage{txfonts}
\usepackage[colorlinks,citecolor=blue,urlcolor=blue]{hyperref}
\usepackage{xcolor}

\begin{document}

   \title{SoFT: Detecting and Tracking Magnetic Structures in the Solar Photosphere}

    \author{M. Berretti \inst{1,2}
            \and
            M. Stangalini \inst{3}
            \and
            S. Mestici \inst{4}
            \and
            D.~B. Jess \inst{5,6}
            \and
            S. Jafarzadeh \inst{7,8}
            \and
            F. Berrilli \inst{2}
            }

    \authorrunning{Berretti et al.}

   \institute{University of Trento,
              Via Calepina 14, 38122 Trento, Italy,
              \email{michele.berretti@unitn.it}
         \and
             University of Rome Tor Vergata, Department of Physics, Via della Ricerca Scientifica 3, 00133 Rome, Italy
         \and 
             ASI Italian Space Agency, Via del Politecnico snc, 00133 Rome, Italy
          \and
            University of Rome "La Sapienza", Department of Physics, P.le A. Moro 5, 00185 Rome, Italy
          \and 
             Astrophysics Research Centre, School of Mathematics and Physics, Queen’s University Belfast, Belfast, BT7 1NN, Northern Ireland, UK
          \and
            Department of Physics and Astronomy, California State University Northridge, Northridge, CA 91330, USA
          \and
             Max Planck Institute for Solar System Research, Justus-von-Liebig-Weg 3, 37077 G\"{o}ttingen, Germany
          \and
          Niels Bohr International Academy, Niels Bohr Institute, Blegdamsvej 17, DK-2100 Copenhagen, Denmark}

   \date{}

  \abstract
   {In this work, we present SoFT: Solar Feature Tracking, a novel feature-tracking tool developed in Python and designed to detect, identify, and track magnetic elements in the solar atmosphere. It relies on a watershed segmentation algorithm to effectively detect magnetic clumps within magnetograms, which are then associated across successive frames to follow the motion of magnetic structures in the photosphere. Here, we study its reliability in detecting and tracking features under different noise conditions starting with real-world data observed with SDO/HMI and followed with simulation data obtained from the Bifrost numerical code to better replicate the movements and shape of actual magnetic structures observed in the Sun's atmosphere within a controlled noise environment.}

   \keywords{Techniques: image processing --
              Methods: data analysis --
              Sun: photosphere
               }

   \maketitle

\section{Introduction}
\label{introduction}

In the past decades, feature tracking algorithms have been used in solar physics to investigate the dynamics of both magnetic and non-magnetic structures within the solar atmosphere for various scientific objectives. Many authors have employed different tracking techniques to study the emergence and diffusion processes of small-scale magnetic structures in the solar photosphere \citep[e.g.][]{nisenson_motions_2003, abramenko_turbulent_2011, giannattasio_diffusion_2013, giannattasio_diffusion_2014, jafarzadeh_migration_2014, lamb_spatial_2014, jafarzadeh_kinematics_2017}, their oscillations through the different layers of the Sun atmosphere \citep[][]{morton_evidence_2013, stangalini_first_2013, 2024arXiv240911553B}, and their statistical properties \citep[][]{parnell_power-law_2009, 2011ApJ...740L..40K, huang_coronal_2012, keys_magnetic_2019}. In addition, feature tracking codes have been employed for the detection of granules in the photosphere to study their statistical properties and dynamics \citep[see][]{november_precise_1988, title_statistical_1989, duvall_time-distance_1997, strous_dynamics_2000, berrilli_2-d_2002}. Finally, these tools have also been used to link the onset of space weather events, such as solar flares, with changes in the magnetic field topology and horizontal velocity flows at different layers of the solar atmosphere \citep[][to name a few]{wang_evolution_1992, anwar_rapid_1993, wang_rapid_2006, wang_observational_2010, higgins_solar_2011, liu_rapid_2012, alvarado-gomez_magneto-acoustic_2012, wang_evolution_2018, 2023ApJS..266...17L}.

Over the years, many tracking codes have been developed by the community that are custom-made for each specific need. In \cite{november_precise_1988} the authors first proposed feature tracking in solar physics to study the proper motion of solar granulation through the introduction of Local Correlation Tracking (LCT). LCT works by detecting displacements that maximise the spatially localised cross-correlation between tracers in two images. In \cite{schuck_local_2005}, the author presented a revised version of the method to account for the magnetic induction equation, improving its reliability in tracking photospheric magnetic footpoints. Finally, in \cite{fisher_flct_2008}, the authors introduced a faster and more efficient method for LCT based on the Fourier transform. An analysis of the limitations of LCT can be found in \cite{potts_reduction_2003}. Alternatively, the Multiple Level Tracking (MLT) algorithm, introduced in \cite{bovelet_new_2001} as an improvement over the previously commonly used Fourier filtering methods \citep{roudier_structure_1986, hirzberger_time_1997}, is a threshold discriminator acting on multiple intensity levels to better define the contours of the detected structures. Since then, it has been modified in \cite{fischer_observations_2019} to study the emergence of a small-scale magnetic flux sheet.

In \cite{welsch_magnetic_2003}, the authors introduced YAFTA (Yet Another Feature Tracking Algorithm), an IDL tracking routine that detects clumps of magnetic pixels with a flux-ranked, downhill labelling algorithm and matches them exploiting overlapping pixels between successive frames. Later, in \cite{deforest_solar_2007}, the SouthWest Automatic Magnetic Identification Routine (SWAMIS) was introduced. The SWAMIS suite is a set of PERL routines that work together to detect and track magnetic features. Unlike YAFTA, it uses hysteresis in both space and time to further boost the confidence in the detected elements. It has been used in a large number of publications related to the dynamics of magnetic field concentrations in the Sun's atmosphere \citep[e.g.][to name a few examples]{meyer_solar_2013, regnier_magnetic_2013, lamb_spatial_2014, lamb_solar_2016, gosic_chromospheric_2018, berretti_unexpected_2024} and can be considered as being at the forefront of tracking algorithms for use with observations of the lower solar atmosphere. 
Furthermore, \cite{2018SoPh..293..123K} employed a method for detecting and tracking linear polarisation features (LPFs) in the solar photosphere, based on an algorithm developed by \cite{2013A&A...549A.116J, 2015ApJ...810...54J}. Their approach involves first identifying contiguous pixels above a signal threshold in linear polarisation maps using a blob analyser algorithm. This algorithm identifies and characterises individual features based on their pixel connectivity and properties. Then, the centre of gravity of each identified feature is tracked in subsequent frames to determine properties such as lifetime and velocity. This technique has been successfully applied to study the dynamics and evolution of LPFs, contributing to our understanding of small-scale magnetic structures in the lower solar atmosphere.

Similar tools, but with different scientific objectives, are CURV \citep{hagenaar_dispersal_1999} and MCAT \citep{parnell_nature_2002}. In \cite{keys_magnetic_2019}, the authors studied the magnetic properties of bright points in the photosphere by tracking bright features using a custom intensity thresholding tracking algorithm and estimated the physical properties of the observed structures. Finally, many other tracking codes have been released that introduce deep learning or novel data manipulation to improve tracking reliability and performance, including the works of \cite{asensio_ramos_deepvel_2017}, \cite{jiang_identifying_2020},  \cite{potts_balltracking_2004}, and \cite{attie_magnetic_2015}. While a thorough review is beyond the scope of the current investigation, further details on the mentioned tracking codes can be accessed via their respective publications.

While many of these tracking codes can be accessed freely and are still functional, they often require paid-for software licences and/or do not make use of modern programming languages that are readily used by the modern community. Here, we present a novel feature tracking tool, SoFT, built in Python with reliable detection and fast associations at its core. Based on a watershed segmentation algorithm, it effectively detects the boundaries of the magnetic structures that are subsequently associated by checking the maximum overlapping features between successive frames. In this work, our aim is to showcase the capabilities of our tracking code with a sample series of magnetograms captured by the Helioseismic and Magnetic Imager \citep[HMI;][]{2012SoPh..275..207S} onboard the Solar Dynamics Observatory \citep[SDO;][]{2012SoPh..275....3P}. Moreover, we provide a study on the reliability of the tool under different noise conditions using simulated magnetic structures obtained from the Bifrost code \citep{gudiksen_stellar_2011}.

\section{Methodology}
The Solar Feature Tracking (SoFT) algorithm can be divided into three main phases. Initially, it detects and identifies magnetic elements according to the input parameters provided by the user. Following detection, the corresponding features in subsequent frames are associated with one another. Finally, in the last phase, it estimates the position, flux, and area of the detected features. Each phase can be parallelized across multiple CPU cores, significantly reducing the code's execution time

\subsection{Detection and Identification: The watershed algorithm}

The detection process starts by masking pixels below a given threshold ($\texttt{l\_thr}$) which is determined based on the noise level in the data. It is common practice to consider a threshold equal to three times the noise level (i.e. 3$\sigma$) to ensure high confidence in the detected structures \citep[e.g., similar to the feature thresholding used in][]{2019ApJ...871..133J}. However, this could potentially result in the loss of many of the fainter magnetic structures. Additional constraints such as the lifetime and size of the detected elements can drastically improve the detection confidence. These constraints can potentially allow one to lower the threshold depending on the signal-to-noise ratio of the instrument, hence increasing sensitivity without losing accuracy. However, it is important to note that lowering the threshold value could result in suboptimal detection of feature boundaries.  We recommend testing different thresholds on the data to determine the optimal value.

For magnetograms, the standard deviation of polarisation signal in continuum position ideally defines this noise level. However, it is not always available, and a reasonable estimate can be obtained from the average value of the standard deviation in quieter regions of the field of view (FoV) of a polarisation image (e.g., a magnetogram). The main drawbacks of this approach are (i) the overestimation of the real polarimetric noise, and (ii) the requirement to perform this operation manually on a single frame rather than automatically for all images in the dataset. 

\begin{figure}[!t]
    \centering
    \includegraphics[]{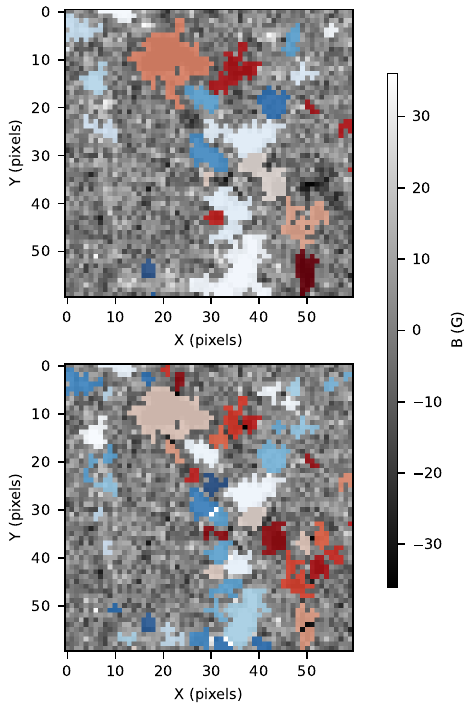}
    \caption{Small crop of the FoV considered in this work. The user parameters chosen are: $\texttt{l\_thr}$ = 12~G, $\texttt{min\_distance}$ = 3~pixels, and $\texttt{m\_size}$ = 4~pixels. The boundaries of the detected features have been highlighted using differently colored contours, respectively blue and red for the positive and negative polarities. The top panel displays the results of the coarse detection method, where nearby magnetic elements of the same polarity tend to be clustered together. The bottom panel, on the other hand, shows the results of the identification step using the finer detection method, which considers each local maxima detected as a single feature. }
    \label{fig:1}
\end{figure}

The next step is to find the local maxima in the masked images. Each peak must be separated by at least $\texttt{min\_distance}$ pixels to prevent fragmentation of a single magnetic structure. The parameter $\texttt{min\_distance}$ can be defined by the user based on the expected size of the magnetic structures to be detected, i.e., through inspection or a-priory knowledge \citep[e.g.,][]{2001ApJ...553..449B, 2008ApJ...684.1469D, 2009MNRAS.397.1852C, 2009A&A...498..289U, 2012ApJ...746..183J, 2014A&A...568A..13R}. 

Then, we computed the Euclidean distance transform (EDT) of all the masked images in the data set. This process replaces every nonzero pixel with the distance to the closest background pixel (i.e. zero-valued). If the pixels are already part of the background, then this distance value is zero. The result of this process is a distance map which provides an approximate gradient field of the image and is used to ensure proper separation of the magnetic structures with the watershed segmentation. 

\begin{figure*}[h!]
    \centering
    \includegraphics[width=\linewidth]{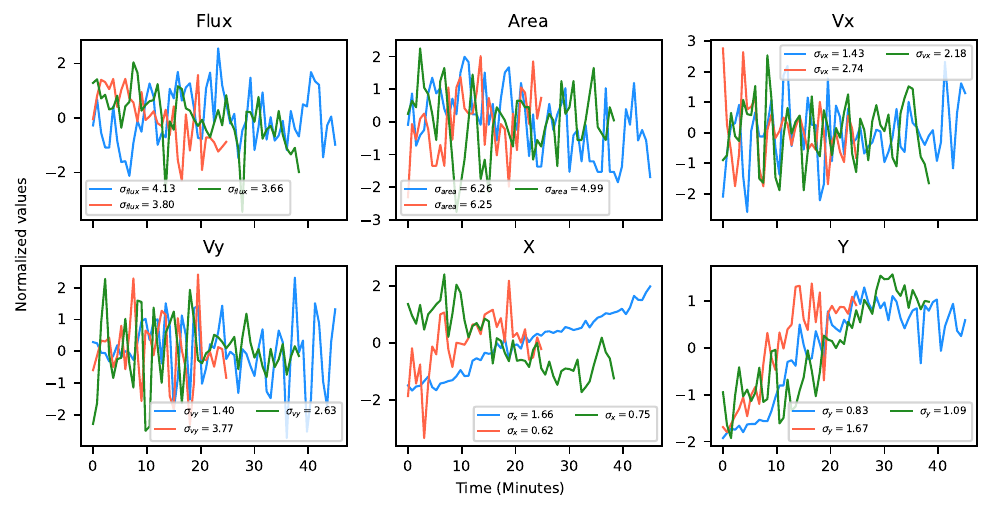}
    \caption{Time series of physical observables obtained from the SoFT tracking code using the finer detection method for three randomly selected elements within the dataset. These plots show the evolution over time of key physical parameters of the magnetic structures, such as magnetic flux, area, horizontal velocity in both $x$ and $y$ directions, and their positions in pixel coordinates. The time series are adjusted to a zero mean and scaled by their respective standard deviations to improve visualisation and the value of $\sigma$ is reported in the legend. Units of measure of the standard deviations are Gauss for flux, pixels$^2$ for area, km/s for velocities, and pixels for positions.}
    \label{fig:2}
\end{figure*}

Finally, the watershed segmentation algorithm is applied to the EDT maps. The previously identified local maxima are used as the initial points from which the watershed algorithm will start flooding the basins, growing regions by following the gradient descent path until a watershed line or a boundary is encountered. The result is a segmented image in which each region corresponds to a distinct magnetic structure. Each detected clump is associated with a unique label and the structures below a given minimum size ($\texttt{m\_size}$), defined by the user, are filtered out. In the top panel of Fig.~\ref{fig:1} we show an example of the elements detected by SoFT in a small FoV in a magnetogram provided by SDO/HMI. Each frame can be assigned to a separate core from the available CPU pool, allowing parallel execution.

In addition to the primary detection procedure, SoFT includes a coarser detection technique that bypasses the distance transform, thus clustering nearby magnetic structures of the same polarity. Its results are shown in the bottom panel of Fig.~\ref{fig:1}. These two methods are referred to as 'fine' and 'coarse', and the user can select one by using the boolean \texttt{separation} flag.

\subsection{Tracking}

Once the features have been identified, they need to be matched from one frame to the next. Following similar tracking codes, such as YAFTA and SWAMIS, we check the overlap between feature \texttt{M} in frame \texttt{n} (denoted as \texttt{M(n)}) and all the features in frame \texttt{n+1} that occupy the same pixels as \texttt{M(n)}. Among these, the feature with the largest overlap is selected as the candidate match for \texttt{M(n+1)}. Once the best guess for \texttt{M(n+1)} is chosen, we repeat the same process backwards and check the overlap between the candidate feature \texttt{M} in frame \texttt{n+1} and all the features in frame n that occupy the same pixels. If the two features, \texttt{M(n)} and \texttt{M(n+1)}, select each other, they are considered matched successfully. This is a successful association and a common label is assigned to both (for convenience, the label of the feature on the frame \texttt{n} is passed to the matched feature on the frame \texttt{n+1}). 

It is important to mention that this tracking method heavily relies on the overlap between same features in subsequent frames. In some cases, an unfortunate combination of temporal cadence of the instrument, its resolution, and the expected size of the features to track, might lead to an improper association, as the features would not overlap. Therefore, it is suggested to review the properties of the instrument against a-priori knowledge of the dynamical and morphological properties of the features to be tracked.

This phase is divided into rounds. Initially, frames are paired and associated two by two, and each pair is condensed into a single unit called a cube (e.g., at the second round each cube will be composed by 4 consistently labeled stacked frames) . In the following rounds, we focus on the last frame of the lower cube and the first frame of the upper cube. The relabeling from this association is then propagated to all the frames within the respective cubes. This method continues to extend progressively until only one cube remains, containing all frames with features associated properly. This process allows the algorithm to spread the workload across all available CPU cores as each cube is assigned to one of the available cores, drastically reducing the running time of the code.

\subsection{Tabulation}

Once the magnetic structures have been successfully detected, identified and associated, we can estimate their physical properties. The positions of the magnetic elements in each frame are obtained by calculating their barycenter, which is obtained by averaging the coordinates of each pixel belonging to a feature weighted by their intensity (i.e., centre of gravity of intensity). This approach provides sub-pixel accuracy on the estimated positions of the detected clumps. The area of each element is measured by counting the pixels within its contour, while the total magnetic flux is measured by summing the intensity of the pixels. Finally, the velocity of the horizontal displacements of the features is estimated by performing the first-order derivative of the positions of the barycenter.

The final output of the SoFT code is a "pandas DataFrame" data structure, exported as a JSON file, providing the following quantities for each of the tracked features:
\begin{itemize}
    \item \textbf{Label:} The unique label assigned to the detected feature during the identification and association process. This label allows for easy reference to specific features in the mask images produced by the tracking code.
    \item \textbf{Lifetime:} The duration of the tracked feature.
    \item \textbf{X and Y:} Arrays containing the positions of the barycentre of the detected structure in each frame.
    \item \textbf{Area:} Array containing the number of pixels inside the contour in each frame.
    \item \textbf{Flux:} Array containing the mean value intensity of pixels inside the contour at each frame.
    \item \textbf{Frames:} The frames in which the selected features appear.
    \item \textbf{$V_{x}$ and $V_{y}$:} Arrays containing the two components of the horizontal velocity.
    \item \textbf{$\sigma_{V_x}$ and $\sigma_{V_y}$:} The standard deviation ($\sigma$) of both components of the horizontal velocity, referring to the entire velocity series.
    \item \textbf{Line-of-Sight Velocity:} If Dopplergrams are available, this provides an array of the mean pixel intensity inside the contour for each Dopplergram frame. 
\end{itemize}

In addition to the DataFrame, the code also saves the masks produced at each step. The total storage required for the output, including both the DataFrame and the masks, is approximately four times the size of the input dataset.

\subsection{Large-scale magnetic structures}

Although SoFT was originally designed to track small-scale magnetic structures in the photosphere with typical equivalent diameters ranging from 300 to 1500 km, it has also been optimised with a dedicated workflow for the detection and tracking of much larger features such as sunspots in the photosphere. Additionally, SoFT can also be used to track features in the upper layers of the Sun's atmosphere (e.g., coronal holes in the corona). The main differences here lie in the detection algorithm, as the other phases remain untouched.

Indeed, when working with larger structures, we want to avoid as much as possible the unwanted splitting of a single feature given by the presence of multiple local maxima in it. In order to do that, the sunspot workflow in SoFT uses a modified version of the detection phase that bypasses the Euclidean distance transform and watershed segmentation, in favour of a much simpler threshold discriminator.

\section{Dataset}

To showcase the results of our tracking suite, we selected magnetograms acquired by SDO/HMI in the Fe~{\sc{i}} 617.3~nm absorption line with a cadence of 45 seconds on 16 April 2020, starting at 00:30 UTC for 45 minutes over a $200\times200$ square arcseconds region at the centre of the solar disk. The different frames have been coregistered with one another in order to obtain a fixed FoV and remove contributions of the solar rotation from the horizontal displacements of the tracked elements. 

We selected the threshold as previously described and opted for $\texttt{l\_thr}= 2\sigma$ above the noise. The value of $\texttt{l\_thr}$ was estimated considering a $5\times5$ pixel$^{2}$ subregion where no magnetic activity is present and was found to be equal to $\approx$6~G. This threshold would imply a confidence of around 95\% for features exceeding this value. However, further constraints on the minimum size of the detected structures (more than 4 pixels) and on their lifetime (more than 4 time steps, i.e. 180~seconds) ensure an higher confidence level. 

To study the performance of the SoFT code under different noise conditions in a controlled environment, we used data from Bifrost simulations. To this end, we have selected magnetic-field maps corresponding to a geometric height of 100 km in the original simulation box. The simulations used in this work reproduce the behaviour of the photosphere below a coronal hole region, representing the quiet Sun. 
This simulation encompasses a three-dimensional space of 768 points in each direction ($x$, $y$, and $z$). Horizontally, this spans 24 Mm, while vertically it covers 16.8 Mm. To mimic the Sun's activity, a slow inflow of horizontal magnetic field is introduced at the bottom boundary of the simulation, positioned 2.5 Mm below the visible solar surface.
We considered a small crop of $1240\times1240$ square kilometres with a pixel size of 31 km and a temporal cadence of 10 seconds. For more details about the simulation, we refer to the simulation run with the identity code \textit{ch024031\_by200bz005}, which is also used and described in \citet{2021SoPh..296...84D} and \citet{2024ApJ...963...10S}.

\section{Results \& Discussion}

In this section, we present the results of the tracking suite under different conditions. First, we show the performance of the algorithm in a normal use case by tracking magnetic structures in magnetograms captured by SDO/HMI. Then, to understand the performance of our tool and estimate its reliability under different noise scenarios, we show the result of the tracking code using simulation data with increasing levels of added noise. 

\subsection{Tracking accuracy in real-world conditions}

Fig.~\ref{fig:1} shows the result of the detection and identification step on a small subset of the FoV considered in this test analysis, with both the coarse (top) and fine (bottom) approaches documented. Both approaches identify the same outer contours for most of the features. However, the fine approach, as intended, further distinguishes and splits the detected elements in order to identify each separate bundle of magnetic field.

For each of the identified magnetic elements in the FoV, the tracking code provides its magnetic flux, area, velocities, and positions in pixel coordinates with respect to the FoV over each frame. In Fig.~\ref{fig:2}, we show the time series obtained for three randomly selected elements in the data set. These time series highlight the reliability and continuity of the tracking code, which is capable of tracking the evolution over time of key physical parameters of the magnetic structures in the lower solar atmosphere.

Finally, in \ref{ap1} we show the statistical distributions of the physical parameters inferred from the tracking suite.

\subsection{Tracking accuracy estimation in controlled conditions}

Establishing a validation pipeline for such a feature detection and tracking tool is a nontrivial task. Indeed, in real use cases, there is no ground truth available to compare the results of the tracking code with. To address this, we tracked features in both modelled and simulated data with no noise. Then, by introducing a controlled amount of noise and repeating the analysis, we can evaluate how the tracking code performs under different conditions. 

First, we tracked the movement of a uniform circular element with a radius of $5$ pixels performing a random walk on a blank canvas of $50\times50$ pixels with an added Gaussian random noise with $\sigma$ equal to 25\% the maximum pixel intensity in the image. The mean difference between the actual trajectory of the ball and the one inferred by the tracking algorithm output is around $0.06 \pm 0.03$~pixels.

\begin{figure}[!t]
    \centering
    \includegraphics[width=7cm]{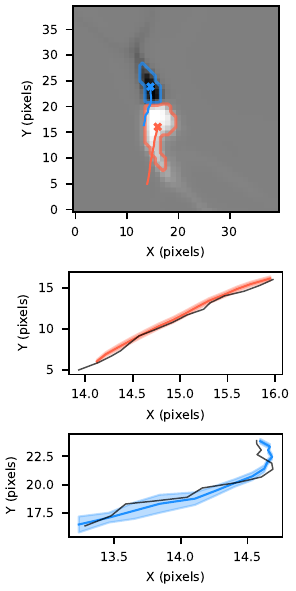}
    \caption{Comparison of the actual and tracked trajectories of a bipolar magnetic structures obtained from Bifrost simulations. The dataset consists of 15 frames. The detected contours and actual trajectories are shown in orange and blue respectively for element 1 and 2. The black line represent the actual trajectories of the two features, obtained by tracking directly in the Bifrost simulated magnetograms. The orange and blue lines show the trajectories tracked by SoFT over $N=100$ iterations with a noise level of 12.5\% of the maximum value of $B$ added to the original image. The shaded areas represent the standard deviation for each frame relative to the average path.}
    \label{fig:3}
\end{figure}

However, a uniform circular element does not accurately represent the shape of a magnetic structure in the solar photosphere. Therefore, we repeat our analysis using a magnetic element derived from simulations. Specifically, we isolated a bipolar magnetic structure generated with the Bifrost code for $15$ frames. Furthermore, to improve the robustness of our results, we repeated the analysis $N=100$ times, producing an ensemble of trajectories shown in the bottom panel of Fig.~\ref{fig:3} in orange and blue, respectively elements 1 and 2 corresponding to the positive and negative polarity of the bipole from here on. These trajectories are compared to the original one, shown in black. It is also worth noting that the amount of noise introduced in the images (a Gaussian random noise with a $\sigma$ equal to 12.5\% the maximum value of the magnetic field strength, $B_{los}$, which is relatively high since we have previously shown that this same quantity is instead equal to about 5\% in the SDO/HMI image sequences) and the size of the elements to track (just barely above the minimum required area) make it the worst-case scenario for the SoFT code. Finally, we performed the same analysis with different percentages for $\sigma$ of added Gaussian random noise with respect to the maximum value of $B_{los}$ and reported the mean deviation from the actual trajectory in Table~\ref{table:1}.

\begin{table}
\caption{Mean deviation from the actual trajectory and the $N=100$ trajectories tracked at different noise levels (D1 ad D2, respectively for element 1 and 2). The table shows the average error in pixels. Columns R1 and R2 show the ratio between the lifetime of the features as inferred by the tracking code and the actual lifetime of the feature.}             % title of Table
\label{table:1}      % is used to refer this table in the text
\centering                          % used for centering table
\begin{tabular}{c c c c c}        % centered columns (4 columns)
\hline\hline                 % inserts double horizontal lines
Noise level & D1 & D2 & R1 & R2 \\    % table heading 
\hline                        % inserts single horizontal line
0.050 & 0.088 & 0.142 & 0.998 & 1.043 \\
0.067 & 0.098 & 0.179 & 1.000 & 0.999 \\
0.083 & 0.151 & 0.217 & 0.995 & 0.991 \\
0.100 & 0.190 & 0.238 & 0.994 & 0.978 \\
0.117 & 0.229 & 0.251 & 0.991 & 0.934 \\
0.133 & 0.302 & 0.302 & 0.976 & 0.946 \\
0.150 & 0.338 & 0.314 & 0.971 & 0.911 \\
0.167 & 0.325 & 0.340 & 0.974 & 0.880 \\
0.183 & 0.381 & 0.363 & 0.972 & 0.885 \\
0.200 & 0.374 & 0.413 & 0.972 & 0.835 \\
\hline                                   %inserts single line
\end{tabular}
\end{table}

\section{Conclusions}
In this work, we presented SoFT, a novel feature tracking suite developed in Python and designed to detect, identify, and track magnetic elements in the solar atmosphere. Built on well-established techniques, such as the watershed segmentation algorithm, and utilising the ease-of-use of Python, it offers a simple and accessible alternative to many of the currently available tracking tools. Through an extensive series of tests, we demonstrate the robustness and reliability of SoFT in different scenarios and noise conditions. 

At first, we studied the performance of the tracking suite on magnetograms captured by SDO/HMI to show how it works in real-world applications. In this case, SoFT has proven capable of detecting, isolating and following the many magnetic features frame by frame, despite the dynamic changes of the magnetic structures further complicating their tracking. In addition, we performed an extensive noise analysis to understand the code's limitations. By tracking both modelled control data and simulated magnetic structures from Bifrost simulations, we were able to establish a baseline for the tracking suite. Even under extremely harsh noise conditions, far above the ones typically observed in actual magnetograms on many different instruments, SoFT demonstrated its robustness by maintaining a high degree of reliability. 

SoFT opens up new possibilities for studying the dynamics of magnetic structures in the solar atmosphere. As it is developed in Python, it is extremely versatile and can be adapted to many different scenarios. Finally, SoFT is a live project, and further iterations of the tracking suite will include the possibility of identifying and tracking granules in continuum images and linking magnetic structures tracked in the photosphere with magnetic features in the upper layers of the solar atmosphere. SoFT is freely available at \url{https://github.com/mib-unitn/SoFT}.

\begin{acknowledgements}
% MB
MB acknowledges that this publication (communication/thesis/article, etc.) was produced while attending the PhD program in Space Science and Technology at the University of Trento, Cycle XXXIX, with the support of a scholarship financed by the Ministerial Decree no. 118 of 2nd March 2023, based on the NRRP - funded by the European Union - NextGenerationEU - Mission 4 "Education and Research", Component 1 "Enhancement of the offer of educational services: from nurseries to universities” - Investment 4.1 “Extension of the number of research doctorates and innovative doctorates for public administration and cultural heritage” - CUP E66E23000110001.
% MS
% SM
S.M. acknowledges the Ph.D. course in Astronomy, Astrophysics and
Space Science of the University of Rome “Sapienza”, University of Rome “Tor Vergata” and Istituto Nazionale di Geofisica e Vulcanologia, Italy.
% DBJ
DBJ acknowledges support from the Leverhulme Trust via the Research Project Grant RPG-2019-371. DBJ wishes to thank the UK Science and Technology Facilities Council (STFC) for the consolidated grants ST/T00021X/1 and ST/X000923/1. DBJ also acknowledges funding from the UK Space Agency via the National Space Technology Programme (grant SSc-009).
% SJ
% FB
% WaLSA
Finally, we wish to acknowledge scientific discussions with the Waves in the Lower Solar Atmosphere (WaLSA; \href{https://www.WaLSA.team}{https://www.WaLSA.team}) team, which has been supported by the Research Council of Norway (project no. 262622), The Royal Society \citep[award no. Hooke18b/SCTM;][]{2021RSPTA.37900169J}, and the International Space Science Institute (ISSI Team~502). 
\end{acknowledgements}

\bibliographystyle{aa} 
\bibliography{main}

\clearpage

\appendix
\section{Statistical characterisation of the dataset}\label{ap1}

\begin{figure}[t!]
    \centering
    \includegraphics[]{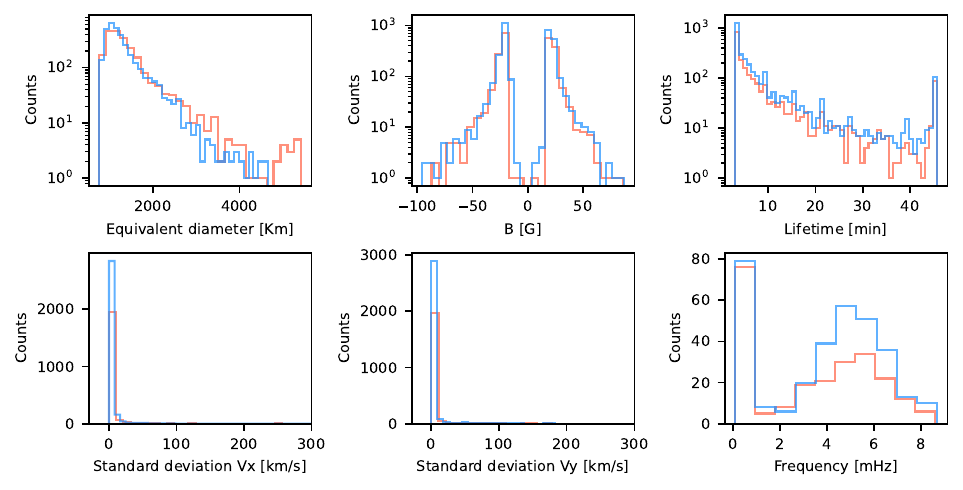}
    \caption{Statistical distributions of the physical properties inferred by the tracking code. The orange distributions correspond to the coarse detection method, while the blue distributions correspond to the fine detection method. From left to right, top to bottom: equivalent diameter, magnetic field, lifetime, standard deviation of $v_x$ and $v_y$, and the distribution of the dominant frequency of $v_x$ following the same procedure of \cite{berretti_unexpected_2024}.}
    \label{fig:ap1}
\end{figure}

In Fig.~\ref{fig:ap1}, we show the statistical distributions of the physical parameters of the magnetic structures detected as inferred by SoFT. Our data set consists of 1 hour of magnetograms captured by SDO/HMI on 16 April 2020, starting at 12:30 UTC for 45 minutes with a cadence of 45~seconds in a $200\times200$ square arcseconds region at the centre of the solar disk. The frames were co-registered to remove the contributions from solar rotation. Magnetic structures are detected using both the fine and coarse approaches. The detection parameters used are the same in both cases except for \texttt{min\_distance}, which is `3' for the fine approach and `5' for the coarse approach. The former resulted in the detection and tracking of 3236 magnetic structures, while the latter led to the detection and tracking of 2174 structures. It is worth noting that the distributions obtained are an example of the magnetic elements considered in this work and are not to be considered representative of the entire population of magnetic structures present in the solar photosphere. 

\end{document}